\documentclass{article}
\input{tcilatex}

\begin{document}

{\Large Some conjectures about the mechanism of poltergeist phenomenon }(%
\footnote{%
An article on poltergeist phenomenon will appear on NeuroQuantology Journal
2008 Vol 6, No 2 (http://www.neuroquantology.com).}){\large \medskip }

P. Brovetto and V. Maxia (\footnote{%
On leave from: Istituto di Fisica Superiore - Universit\`{a} di Cagliari -
Italy.
\par
Email: \ pbrovetto@gmail.com \ \ \ })

\textbf{Summary - }Poltergeist accounts concern at least four kinds of
strange spontaneous manifestations, such as burning of materials, failures
of electric equipments, rapping noises and movements of objects. A simple
analysis of phenomenology of these disturbances shows that they might have a
common origin, that is, a reduction in strength of molecular bonds due to an
enhancement in polarization of vacuum which decreases the actual electron
charge. Arguments based on Prigogine' nonequilibrium thermodynamics are
proposed, which show how transformations in brain of some pubescent children
or young women might be the cause of these effects.


\medskip

\textbf{1 - Introduction}

Accounts about poltergeist are common to different ages and different
countries around the world. They concern certain strange spontaneous
phenomena, such as rappings, burning of fabrics, failure of electric
equipments, movement of objects, for which there is no acceptable
explanation. Often these disturbances occur in the neighbourhood of one
person, generally a pubescent child or a young woman. Believers in spiritism
interpret these phenomena as evidence of the presence of supernatural
spirits. Misbelievers are inclined to deny their existence. Sometimes this
position is justified by the little reliability of persons which witnessed
the event. They indeed frequently embroider their accounts with fanciful
details.

Nevertheless, in the last few decades some poltergeist occurences have been
reported in which qualified persons attended the fenomena. Let us quote the
case of Canneto of Caronia, a village near Messina, in which fabrics,
electric lines and equipments of various kinds burned unexpectedly in
presence of police and fireman officers $\left[ 1\right] $. A similar
episode is that of Falciano of Massico, a village near Caserta, in which
curtains, blankets and shirts caught fire without a reason. These events
took place only when a 11-year old girl, Isabella, was in the neighbourhood $%
\left[ 2\right] $. In year 1947, one of us (P.B.) attended to a different
kind of poltergeist event, that is, rapping disturbances taking place in a
cinema in Turin (See Appendix A). As for the movement of objects, it is
worth quote the case of wine bottles pushed down from the rack, which was
attended by the renowned anthropologist Cesare Lombroso in the distant year
1900 $\left[ 3\right] $. But, there is a case which was validated, besides
by public officers, by physicists from the Max Planck Institute. It took
place in years 1967-68 in a lawyer's office in Rosenheim (Bavaria). Various
heavy damages were caused to office machinery and to electric and telephone
lines. The focus of these events seemed to be Annemarie, a nineteen-year old
employee. The disturbances ceased when she left the office, although
witnesses claimed further events took place in her new office. One of
physicists of Max Planck Institute, Dr. F. Karger, said "What we saw in the
Rosenheim case could be 100\% shown not to be explainable by known physics" $%
\left[ 4\right] $.\medskip

\textbf{2 - The phenomenology of poltergeist disturbances.}

The first step in order to find a plausible\ meaning to the previous jumble
of disparate events is that of analyzing them separately on the ground of
known physics. Since light fire is a quite current job, let consider first
the case of burning materials.

\textit{a)} Burning materials. - Burn of paper, cotton or woollen fabrics is
one of the most frequent poltergeist events. Actually, combustion is an
oxidation reaction involving atmospheric oxygen, ruled by the velocity law
of all chemical reactions, that is, the Arrhenius' law $\left[ 5\right] $.
Dependence on temperature is characterized by the heat of activation $W_{0}$%
, that is, the energy required to remove bonds in the reacting molecules so
allowing the formation of oxygen bonds in the combustion products. By
letting $A$ be the Arrhenius' pre-exponential factor, the combustion
velocity $v\left( T\right) $ is given by%
\[
v\left( T\right) =A\exp \left( -\frac{W_{0}}{RT}\right) ;\quad \left(
T_{0}\simeq 300\text{ K}\right) \left. 
\begin{array}{c}
T=2T_{0}\simeq 300^{\circ }\text{C} \\ 
T=3T_{0}\simeq 600^{\circ }\text{C}%
\end{array}%
\right\} ignition, 
\]%
where $R=1.98$ cal mole$^{-1}$ K$^{-1}$. According to this equation, when $T$
is enhanced from the room value $T_{0}\simeq 300$ K, ignitions begin to take
place with increasing velocities. For $T=2T_{0}$ ($\simeq $ $300^{\circ }$%
C), various materials already catch fire, but for $T=3T_{0}$ ($\simeq $ $%
600^{\circ }$C) practically all materials ignite. On this ground, to account
for room temperature ignition we are forced to assume that in poltergeist
events an enigmatic agent succeeds in decrese activation energy $W_{0}$\ in
such a way that the Arrhenius' law can be rewritten in the form%
\[
v\left( W\right) =A\exp \left( -\frac{W}{RT_{0}}\right) ;\quad \left(
T_{0}\simeq 300\text{ K}\right) \left. 
\begin{array}{c}
W=W_{0}/2 \\ 
W=W_{0}/3%
\end{array}%
\right\} ignition. 
\]%
This amounts to admit that the poltergeist burnings are originated by a
decrease in the molecular bond strengths.

\textit{b)}\textbf{\ }Failures of electric equipments. - Indeed, the
previous interpretation easily explains also the failure of electric
equipments, sometimes observed in poltergeist events. Actually, insulators
utilized in these equipments are made up of rubber or plastic in which
chains of carbon atoms are linked by covalent bonds. Weakening in bonding
energy reduces the dielectric strength of insulators thus originating their
electric breakdown.

\textit{c)}\textbf{\ }Rappings. - The most intriguing occurrence is perhaps
that of the drumming noises. It could happen that the decrease in bonding
energy is great enough that some oxygen molecules in the air are split in
two atoms. These molecules own three degrees of freedom due to their
movement in space, two degrees due to rotation around two axes mutually
orthogonal and orthogonal to the line joining the atoms and one degree due
to anharmonic oscillation of atoms along this line. In total six degrees of
freedom. Two oxygen atoms also own six degrees due to their independent
movements in space. Owing to the kinetic energy equipartition theorem which
allots energy $kT/2$ to each degree of freedom, the energy owned by these
degrees is always $6kT/2$ ($k=1.38\times 10^{-23}\unit{J}\unit{K}^{-1}$) $%
\left[ 6\right] $. So, no effect on kinetic energy is expected due to the
molecule splitting, while, on the contrary, pressure of oxygen is doubled. A
fast increase in pressure could originate a burst and a succession of bursts
the drumming noise. Obviously, the same effect is allowed for nitrogen
molecules.

\textit{d)}\textbf{\ }Moving objects. - As for the strange movements of
objects, they can be explained by slow increases in pressure that put
strengths on the objects without originating bursts.

It follows that all disturbances could be ascribed to a weakening in
molecular bond energies. It is to be pointed out, however, that in
disturbances at items \textit{a)}\textbf{\ }and\textbf{\ }\textit{b)}\textbf{%
\ }the energies involved are the heat of combustion of the burned materials
and the energy supplied by the power lines, respectively. The poltergeist
agent acts only as a trigger. On the contrary, for disturbances at items 
\textit{c)} and \textit{d)}\textbf{\ }no external source is active so that
energy actually derives from the poltergeist itself. In this case, its
effect is significant, since bond-dissociation energies are as high as 118
kcal mole$^{-1}$ and 225 kcal mole$^{-1}$ for $O_{2}$ and $N_{2}$,
respectively.\medskip

\textbf{3 - The effect of quantum fluctuations of vacuum on energy of
molecular bonds}.

The next step, to get a rational interpretation of poltergeist phenomenon,
is that of provide a physical basis to the "enigmatic agent" before quoted.
For this purpose, it is to be considered that strength of molecular bonds
depends on the actual value of electron charge $e=1.6\cdot 10^{-19}\unit{C}$%
. It, however, is not an absolute constant, as light velocity $c$ and
Planck' constant $h$. Charge $e$ is a "renormalized" constant. Indeed, owing
to quantum fluctuations due to the energy-time indeterminacy principle $%
\delta w\ \delta t\simeq h$, virtual electron-positron pairs continuously
pop out of the vacuum lasting for a time $\delta t\simeq
h/2m_{e}c^{2}=4\cdot 10^{-21}$s, $m_{e}$ standing for the electron mass.
Consequently, vacuum behaves like a polarizable dielectric which has the
effect of reducing all charges by a constant amount $\left[ 7\right] $. So
charge $e$ is smaller than charge $e_{0}$ of the bare electron, that is,
electron without the effect of vacuum polarization. On this basis, and
taking into account that poltergeist occurences, as burns, electric failures
and noises, are always localized in space, we ascribe poltergeist to the
appearance of bubbles or plumes in which vacuum polarization is
substantially enhanced, so reducing $e$ and consequently the chemical bond
strengths. Of course, explain how these perturbations are originated is a
rather exacting thing. It is the subject of the next section.

It is worth to point out that the possibility of influence quantum
fluctuations of vacuum by introducing external constraints was considered by
the Dutch physicist H. Casimir since 1948 $\left[ 8,9\right] $. It showed
that two parallel conducting plates placed at a small distance attract each
other as a consequence of modification of vacuum state in the space between
the plates. Successful conclusive experiments on this matter have been
recently performed $\left[ 10,11\right] $. Although Casimir' effect has
anything to do with the present problem, it is interesting because it proves
that vacuum is indeed modifiable.\medskip

\textbf{4 - A possible cause of the enhancement in vacuum polarization. }

As said before, poltergeist disturbances often occur in the neighbourhood of
a pubescent child or a young woman. These fellows constitute a considerable
part of people, so that it is likely that some of them are always present in
the nearness of places where poltergeist disturbances happen. Puberty is a
modification of the child body which involves various organs, chiefly the
brain. To understand how this modification might in rare cases act on vacuum
polarization, it is worthy to recall some topics in thermodynamics which, in
our opinion, have relation with the matter at issue.

In 1871 J. C. Maxwell investigated the possibility of escape the second law
of thermodynamics, that is, of obtain work from thermal energy without
utilizing a temperature drop. Maxwell considered a vessel filled of an ideal
gas, divided into two equal sections $A$ and $B$ by means of a screen in
which a small window was opened. A being, the devil, able to follow the
movement of individual molecules operates the window in such a way to allow
molecules to pass only from section $A$ to section $B$. After a determinate
time, all molecules are gathered in section $B$ so that, if gas returns in
section $A$ flowing through a turbine, work might be obtained. In reality,
the removal of gas from section $A$ to section $B$ entails a decrease in
entropy. At the beginning,$\ $when the window is opened, $N$ molecules of
gas are distributed at random between sections $A$ and $B$. By letting $%
P_{A} $ and $P_{B}$ be the thermodynamic probabilities for placing one
molecule in section $A$ or $B$, respectively, initial probability for
placing $N$ molecules in the whole vessel is $\left( P_{A}P_{B}\right) ^{N}$%
. Likewise, final probability is $P_{B}^{N}$. Taking into account Boltzmann'
law on entropy, we have%
\[
S_{B}-S_{Ran}=k\log P_{B}^{N}-k\log \left( P_{A}P_{B}\right) ^{N}=-Nk\log
P_{A}, 
\]%
$S_{B}$ and $S_{Ran}$\ standing for final and initial entropy, respectively.
But this entropy decrease is not allowed by the second law of
thermodynamics. It has been shown, in fact, that, in order to gain the
information for control the window, the devil must increase the entropy of
gas precisely of the same amount of the decrease obtained (often called
negentropy) $\left[ 12\right] $. Consequently, no free decrease of entropy
is possible and the Maxwell' devil is exorcized.

Let us turn now to a different subject, that is, the chirality
(mirror-symmetry) of proteins. Proteins consist of chains of thousands of
amino acids in which one carbon atom $C^{\ast }$, referred to as asymmetric
carbon, is joined by four tetrahedral bonds to a carboxil $COOH$, an amino
group $NH_{2}$, a hydrogen atom and a radical $R$. Protein chains are formed
by removal of water molecules from carboxils and amino groups. Owing to
asymmetry of carbon $C^{\ast }$, two mirror forms, right ($R$) and left ($L$%
), of amino acids are possible as shown in the following figure%
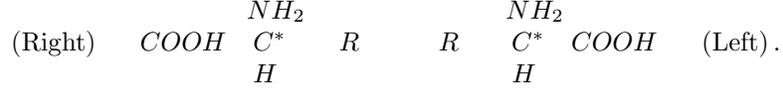
\[
\left( \text{Right}\right) \quad 
\begin{array}{c}
\qquad NH_{2} \\ 
COOH\quad C^{\ast }\qquad R \\ 
\qquad H\,\,\ \,%
\end{array}%
\quad 
\begin{array}{c}
NH_{2} \\ 
\quad R\qquad C^{\ast }\quad COOH \\ 
H\,\,\,\ 
\end{array}%
\quad \left( \text{Left}\right) . 
\]%
(\textit{To read this figure, consider }$C^{\ast }$ \textit{placed on leaf
plane,} $NH_{2}$ \textit{and} $H$ \textit{behind,} $COOH$ \textit{and} $R$ 
\textit{before the plane}). This state of affairs mimics that of the
distribution of molecules in sections $A$ and $B$ of the previous Maxwell'
problem. Indeed, like a molecule can be placed in section $A$ or $B$, bonds
of carbon $C^{\ast }$ can be set in form $R$ or $L$. Place all molecules in
section $B$ or, likewise, set all bonds in form $L$ corresponds to minimum
entropy. By letting $N_{b}$ be the number of different biological amino
acids ($N_{b}=20$), entropy of a protein including at random $N$ amino acids
of forms $R$ and $L$ is

\[
S_{Ran}=k\log \left( 2N_{b}\right) ^{N}, 
\]%
since there are $2N_{b}$ choices for each amino acid in the chain.
Correspondingly, entropy of a protein including only $L$ (or $R$) amino
acids is%
\[
S_{L}\equiv S_{R}=k\log \left( N_{b}\right) ^{N}. 
\]%
It follows that an entropy decrease%
\[
S_{L}-S_{Ran}=-kN\log 2 
\]%
is obtained when transforming a random protein in a $L$ (or $R$ ) chain.
But, as before the second law rules out this way of minimizing entropy.
Surprisingly, all biological proteins are constituted only of $L$ amino
acids. It happens like a devil cleverer than Maxwell' devil operating on
protein chirality succeds in violating the second law!

To solve this conundrum, it is right resort to Prigogine' nonequilibrium
thermodynamics of open systems, that is, systems not insulated from their
environment $\left[ 13,14,15\right] $. It has been shown that in these
systems processes not too far from equilibrium keep production of entropy at
the minimum rate. These processes are stationary, that is, systems remain
unchanged in time. On the contrary, in case of high irreversibility, systems
become unstable and fluctuations are originated which decrease the system
entropy until the so-called dissipative structures appear. This phenomenon
is referred to as order by means of fluctuations or self-organization in
nonequilibrium systems. Of course, this entails that an amount of entropy, $%
S_{Env}$, greater than the decrease produced is thrown in the system
environment by means of fluctuations, in such a way that total entropy, of
system plus its environment, increases as required by the second law. That
is, if%
\[
\left( \frac{dS}{dt}\right) _{Sys}<0, 
\]%
then%
\[
\left( \frac{dS}{dt}\right) _{Env}>>0,\quad \quad \left( \frac{dS}{dt}%
\right) _{Sys}+\left( \frac{dS}{dt}\right) _{Env}>0. 
\]%
Correspondingly, the first law requires that energy is turned out in the
system environment in cases in which self-oganization decreases the energy
of the system.

The theory is rather complex and we do not dwell on this matter. Among the
various examples of dissipative structures providing evidence about the
fitness of this theory, we limit ourselves to quote the oscillating reaction
of Belousov and Zhabotinskij $\left[ 16,17\right] $. It consists in the
oxidation of the malonic acid $COOH.CH_{2}.COOH$ in watery solution by means
of potassium bromate in presence of $Ce^{+4}$ which acts as an oxidation
gauge. In a flat vessel (Petri dish), the dissipative structure appears as a
nice pattern of spiral waves in which concentration of reagents is different
from that of the unperturbed solution. In Appendix B, a practical recipe is
given for producing the reaction dealt with.

The chiral asymmetry of proteins is a clear signature that the organism of
human beings is a dissipative structure memorized in DNA, consequence of a
slow self-organization process implemented in the course of geological ages (%
\footnote{%
Evidence about the basic role of chirality in control living organism
structure is provided by the unfortunate episode of thalidomide drug. It was
patented in 1957 as a sedative and hypnotic. Thalidomide is a glutamic acid
derivative showing an asymmetric carbon in the glutarimide ring. This
chemical is a random mixture of $R$ and $L$ $\ $forms. When it was
administered to expectant mothers the embryonic development was severely
damaged by mismatching between drug and protein chiralities. \ }). However,
at birth and in the following years the brain apparatus which rules
sexuality is not ready. On this basis, the most conservative hypothesis is
that a special self-organization process takes place in brain during years
of puberty. This process, very fast as compared with that creating the human
being organism, should activate the network of sexual neurons. Electrons in
neuron molecules should be engaged in entropy-decreasing fluctuations
leading to a dissipative structure. But brain is imbued with the vacuum
distribution of electron-positron pairs which fills the space. Consequently,
these fluctuations should throw excess entropy $S_{Env}$ into surrounndig
space so enhancing there the density of pairs, that' to say, the vacuum
polarization. The enhanced polarization bubbles hypothized in Section 3\
should be originated in this way. In rare cases, perhaps much less than one
in a million, there are phases in which fluctuations are produced with a
rate great enough that\ the enhanced polarization originates poltergeist
phenomena. This process obviously concerns a limited environment of brain,
of size perhaps not exceeding some meters across.\medskip

\textbf{5 - Final remarks}

Altogether, the previous arguments can be summarized as follows. A decrease
in entropy (creation of order) in brain of pubescent people throws a greater
amount of entropy (disorder) into the brain environment, which, in
exceptional cases, originates poltergeist disturbances. In practice,
poltergeist is interpreted as a by-product of the entropy increase $\left(
dS/dt\right) _{Env}$ expected in consequence of the second law. This
interpretation is based on two sound achievements of the past century
physics, that is, quantum electrodynamics of vacuum and nonequilibrium
thermodynamics.

To develop further the proposed interpretation, topics about thermodynamics
of perturbations in vacuum and of transformations in brain should be
considered. We leave this out but consider the important topic of
poltergeist energy. In disturbances at issues \textit{a)} and \textit{b)} of
Section 2, energies involved, as already pointed out, seem negligible. On
the contrary, disturbances at issues \textit{c)} and \textit{d)}\textbf{\ }%
require\textbf{\ }perhaps even\textbf{\ }some hundred joules. Exceptional
transformations in the human brain might supply this amount of energy, but
certainly cost in energy is a point which limits the frequency of
poltergeist phenomena.

We point out, finally, that the opinion that poltergeist is connected to
anomalies in the vacuum state is not new. Actually, physicists D. Radin,
head of electrical engineering department at Duke University, assumes that
poltergeist movements are repulsive versions of the Casimir effect that can
put pressures on objects (see item \textit{d) }in Section\textit{\ }2) $%
\left[ 18\right] $. Indeed, while the attractive Casimir effect is due to a
decrease of vacuum polarization, its repulsive version should be ascribed to
an enhanced polarization. Consequently, this interpretation agrees with that
proposed in Section 3.\medskip

\begin{center}
\textbf{Appendix A}
\end{center}

\textbf{A poltergeist event in Turin (Spring 1947).}

The event I dwell on took place in a cinema in the Turin centre (i.e. cinema
"Romano" near to "Castello" place). I watched the first afternoon film
presented in this cinema in company with a girl friend. I and my friend took
place rather far from the screen, over half the room length. About ten
minutes after the beginning of the projection, the room was still almost
empty. Nearly one dozen people were gathered near the screen, plus few
lonely people scattered at various distances from the screen.

Unexpectedly, a loud\ drumming noise at a distance less than ten seats
beside my seat came to disturb the projection. It looked like anyone
hammered rapidly the wooden back of its front seat. But nobody was near the
place of the noise origin. After perhaps four seconds, the noise began move
swiftly toward the screen, then it came back and so on several times. This
troubling noise gave rise to energetic protests of spectators, especially
those seated near the screen. These persons, evidently boys, came out with
howls and bad words to induce the disturber to stop. All this uproar ceased
abruptly after less than five minutes. Then the film projection continued
undisturbed while other spectators progressively reached the room. The whole
event repeated itself other two times after intervals of less than one hour.
But, in the meantime all seats in the room were filled. Consequently, the
wooden backs of seats, damped by the occupant spectators, lacked the
resonant capabilities of vacant seats. This not least because persons wore
coats owing to the fresh season. Evidently the drumming noise was originated
somewhere else.

Probably most spectators perceived the strange event as a successful joke
organized by some carefree persons. I and my friend, incapable of explaining
the drumming noise trick, were bewildered but satisfied with the unexpected
free show.\medskip

\begin{center}
\textbf{Appendix B}
\end{center}

\textbf{The Belousov - Zhabotinskij reaction.}

In order to allow a quick start of the reaction, malonic acid is substituted
by bromomalonic acid $COOH.CBrH.COOH$ (BMA) according to the following
recipe:

BMA\ (M.W.182.96)\qquad\ \qquad 0.57\ gr

$KBrO_{3}\ $(M.W.167.01)\qquad \qquad 1.25 gr

$H_{2}SO_{4}\ $2 M\qquad \qquad \qquad \qquad\ \ \ 4.65 ml

Ferroin 0.025 M\qquad \qquad \qquad\ \ \ \ \ 3 ml

Dilute with water to 25 ml. Ferroin oxidation gauge replaces Ce$^{+4}$ ions.

Preparation of BMA: dissolve10.4 gr of malonic acid in 75 ml of ether cooled
with ice, add slowly 16 gr of $Br_{2}$. Evaporate the ether and keep 12
hours in dessicator with $KOH$ under vacuum.\medskip

\begin{center}
\textbf{References}
\end{center}

$\left[ 1\right] $ La STAMPA (Turin) - February 11, 12, 2004.

$\left[ 2\right] $ Ibidem - October 1, 1988.

$\left[ 3\right] $ Ibidem - November 19, 1900.

$\left[ 4\right] $ R. E. Guiley, "Encyclopedia of the Strange, Mystical \&
Unexplained" (Gramercy Books, New York ,1991).

$\left[ 5\right] $ R. H. Fowler "Statistical Mechanics" (Cambridge
University Press, $1955$) pag. 702.

$\left[ 6\right] $ Ibidem pag. $60-63$; R. H. Fowler and E. A. Guggenheim
"Statistical Thermodynamics" (Cambridge University Press, 1949) pag. $%
123-124 $.

$\left[ 7\right] $ W. Heitler "The Quantum Theory of Radiation" (Oxford at
the Clarendon Press, 1970) \S\ 32.

$\left[ 8\right] $ H.G.B. Casimir, Proc. Kon. Ned. Akad. Wet. \textbf{51},
49 (1948).

$\left[ 9\right] $ G. Plunien, B. Muller and W. Greiner "The Casimir effect"
Phys. Report \textbf{134}, 87 (1986).

$\left[ 10\right] $ S. K. Lamoreaux, Phys. Rev. Lett. \textbf{78}, 5 (1997).

$\left[ 11\right] $ G. Bressi, G. Carugno. R. Onofrio and G. Ruoso, Phys.
Rev. Lett. \textbf{88}, 41804 (2002).

$\left[ 12\right] $L. Brillouin "La Science\ et la Theorie\ de\
l'Information" Ch. XIII (Masson, Paris 1959).

$\left[ 13\right] $ I. Prigogine "Introduction to Thermodinamics of
Irreversible Processes" (John Wiley, N.Y. 1986).

$\left[ 14\right] $ G. Nicolis and I. Prigogine "Self-Organization in
Nonequilibrium Systems" (John Wiley, N.Y. 1977).

$\left[ 15\right] $ I. Prigogine and D. Kondepudi "Thermodynamique. Des
moteurs thermiques aux structures dissipatives" Ch. 19 (Editions Odile
Jacob, Paris 1999).

$\left[ 16\right] $ B.P. Belousov, "Sborn referat. radiat. Meditisin za
1958" (Megdiz, Moskva1959) pag. 145.

$\left[ 17\right] $ A.M. Zhabotinskij, Biofyzika \textbf{2,} 306 (1964).

$\left[ 18\right] $ http://www.answers.com/topic/poltergeist, item "Caused
by physical forces" (10/8/2007) pag. 4.

\end{document}